\documentstyle[aps]{revtex}
\begin{document}
\title{Electron-phonon interaction in CdTe/CdMnTe/CdMgTe quantum wells}
\author{X. F. Wang}
\address{Centro Brasileiro de Pesquisas F\'{\i}sicas\\
Rua Xavier Sigaud 150\\
22290-180 Rio de Janeiro, RJ, Brazil and\\
Instituto de F\'{\i}sica, Universidade do Estado do Rio de Janeiro\\
Rua S\~{a}o Francisco Xavier 524\\
20550-013 Rio de Janeiro, RJ, Brazil}
\author{I. C. da Cunha Lima}
\address{Instituto de F\'{\i}sica, Universidade do Estado do Rio de Janeiro\\
Rua S\~{a}o Francisco Xavier 524\\
20550-013 Rio de Janeiro, RJ, Brazil}
\author{A. Troper}
\address{Centro Brasileiro de Pesquisas F\'{\i}sicas\\
Rua Xavier Sigaud 150\\
22290-180 Rio de Janeiro, RJ, Brazil and\\
Instituto de F\'{\i}sica, Universidade do Estado do Rio de Janeiro\\
Rua S\~{a}o Francisco Xavier 524\\
20550-013 Rio de Janeiro, RJ, Brazil}
\date{Draft ---\today }
\maketitle
\newpage
\begin{abstract}
The optic vibrational (confined and interface) modes and the electron-phonon
and hole-phonon interactions are obtained, in dielectric continuum model,
for two diluted magnetic semiconductor structures: a single well of $%
Cd_{1-x}Mn_xTe/Cd_{1-y}Mg_yTe$ with the magnetic ions in the well, and a
double well of $CdTe/Cd_{1-x}Mn_xTe$, in which a thin layer of the magnetic
material is grown in the middle of the structure. The scattering rates for
the intra- and the inter-subband transitions are obtained for electrons and
holes.
\end{abstract}

\section{Introduction}

During the recent years there has been increasing interest and considerable
experimental and theoretical activity focused on the semimagnetic
semiconductors, also called diluted magnetic semiconductors (DMS) \cite
{jain,dms1}. These compounds have some unique properties leading to their
potential use in a wide range of opto-electronic applications. Since 1977,
when Kamarov {\it et al} \cite{kama} first reported the giant enhancement of
magnetic-optical effects in $Cd_{1-x}Mn_xTe$, much effort has been directed
towards the understanding of the physics underlying the unusual phenomena
associated with these special semiconductors. As its non-magnetic
counterpart, $Cd_{1-y}Mg_yTe$ has attracted much attention \cite{reus,sire},
due to their similarities in crystalline and electronic properties, and, as
a consequence, the feasibility of the fabrication of structures such as
quantum wells, quantum wires and dots \cite{illi}. For this reason, with the
development of the technology of the molecular beam epitaxy, microstructures
of DMS/non-magnetic semiconductors, built with the constituents $CdTe$, $%
Cd_{1-x}Mn_xTe$, $Cd_{1-y}Mg_yTe$, and the quaternary alloy $%
Cd_{1-x-y}Mn_xMg_yTe$, have been grown, and their magneto-optic properties
have been largely explored. For instance, in structures built with a
non-magnetic quantum well ($CdTe$) surrounded by DMS barriers, the magnetic
tuning of the barrier potential induces important changes in the energy of
the confined states, which appear as large Zeeman effects of the excitons in
the quantum well \cite{wasi,gaj}. Structures where the DMS layer ($CdMn$)$Te$
is surrounded by non-magnetic material ($CdTe$ or $Cd_{1-y}Mg_yTe$) are also
greatly interesting \cite{buss,test,lema}. Therefore, it is worthwhile to
understand in greater details the vibrational modes and the electron-phonon
interaction in structures with these II-VI DMS. In particular, for the
magneto-optic properties, as well as for the possible magnetic order, holes
play an important role. We will give, in consequence, special attention to
the hole-phonon interaction, considered here as being
described by the same Fr\"olich-like electron-phonon potential, calculated
with the proper hole parameters. The lattice dynamics of the bulk $%
Cd_{1-y}Mn_yTe $ has been studied in detail, both experimentally and
theoretically \cite{picq,venu,pete}. Basically, optical phonons always play
an important role in determining many physical properties in those
materials. Due to the confinement, Faraday rotation \cite{buss}, Kerr effect 
\cite{test}, and RKKY interaction \cite{dietl,bose}, as well as the
optical vibrational modes, and the electron-phonon Fr\"{o}hlich interaction 
may change greatly. At present, the effect of phonon confinement and its 
consequences on the electron-phonon interaction in polar
semiconductor heterostructures seems to be reasonably understood
\cite{mori,george}. In this paper, we will study the confinement
effect on the carrier-optical-phonon interaction in two typical
microstructures with a DMS thin film surrounded by non-magnetic material: a $%
CdTe/CdMnTe$ double well structure (in which case the $CdMnTe$ layer acts as
a barrier), and a $CdMnTe/CdMgTe$ single quantum well (in which case the $%
CdMnTe$ layer is programmed to work as a well).

\section{Electron-optical-phonon interaction}

According to the dielectric continuum model \cite{mori,kim}, the
optical-phonon modes of small wavevector in a microstructure can be obtained
by describing them in several layers with different dielectric functions. In
each layer the dielectric function is expressed by the generalized
Lyddane-Sachs-Teller relation, given in binary semiconductors by: 
\begin{equation}
\varepsilon _1(\omega )=\varepsilon _{1\infty } \frac{(\omega ^2-\omega _{%
\text{L1}}^2)} {(\omega ^2-\omega _{\text{T1}}^2)},  \label{diel1}
\end{equation}
and, in ternary semiconductors: 
\begin{equation}
\varepsilon _2(\omega )=\varepsilon _{2\infty } \frac{(\omega ^2-\omega _{%
\text{L2}\alpha }^2)} {(\omega ^2-\omega _{\text{T2}\alpha }^2)}\frac{%
(\omega^2 -\omega _{\text{L2}\beta }^2)}{(\omega ^2-\omega _{\text{T2}\beta
}^2)}.  \label{diel2}
\end{equation}
Here $\omega $ is the phonon frequency. The layers of different materials
are denoted by numerical subscripts $1$ or $2$, the subscript L (T)
indicating the bulk longitudinal (transverse) mode, and the subscripts $%
\alpha $, $\beta $, etc, indicating different phonon branches in the bulk
ternary alloy. $\varepsilon _{i\infty }$ is the high frequency dielectric
constant of material $i=1$, or $2$. The optical phonon modes satisfy the
classical electrostatic equations in each layer:

\begin{eqnarray}
\nabla ^2\Phi ({\bf r}) &=&\frac 1{\epsilon _0}\nabla \cdot {\bf P}({\bf r})
\\
{\bf P}({\bf r}) &=&\epsilon _0\chi _i(\omega ){\bf E}({\bf r}) \\
{\bf E}({\bf r}) &=&-\nabla \Phi ({\bf r}),
\end{eqnarray}
where ${\bf P}({\bf r})$ is the polarization field, ${\bf E}({\bf r})$ is
the electric field, and $\chi _i(\omega )=\varepsilon _i(\omega )-1$ is the
dielectric susceptibility in lay $i$. The electric potential $\Phi ({\bf r})$
generated by the phonon modes, which interact with electrons through the
Fr\"{o}hlich coupling, and the normal component of electric displacement $%
\varepsilon _i(\omega )d\Phi /dz$, are continuous at each interface \cite
{mori}. The phonon modes are divided into two kinds. One kind is formed by
the bulk-like modes, including the confined modes and the half space modes,
the other kind is formed by the interface modes. The electric field induced
by each phonon mode, is normalized {\it via} the integral \cite{kim} : 
\begin{equation}
\int d{\bf r}\left| \nabla \Phi ({\bf r})\right| ^2\beta _i=\frac{\hbar
\omega }{2\varepsilon _0},
\end{equation}
with $\beta _i=\frac \omega 2\frac{\partial \varepsilon _i}{\partial \omega}$%
. In planar microstructures, the electric potential has the form of $\Phi _{%
\vec{q}}({\bf r})= \frac 1{\sqrt{S}}\phi _{\vec{q}}(z)e^{i{\vec{q}}\cdot 
{\bf \vec{\rho}}}$, with ${\vec{q}}$ being the phonon mode wavevector in the 
$x-y$ plane. $\phi _{\vec{q}}(z)$ is the phonon potential along the $z-$
direction (growth direction), and its form depends on the structure
architecture, and on the phonon mode.

The calculations are performed for the two structures. First, we consider a
symmetric $CdTe$/$Cd_{1-x}Mn_xTe$ double well, as sketched in Fig. \ref
{fig01}: two miniwells of width $L$ are separated by a $Cd_{1-x}Mn_xTe$
barrier of width $d$. The potential of the barrier for electrons (holes) is
the mismatch between the conduction (valence) band edges of $CdTe$ and $%
Cd_{1-x}Mn_xTe$. The electron (hole) wavefunction can be obtained by solving
the Sch\"{o}dinger equation in this structure assuming parabolic energy
bands \cite{wang1}. The second structure to be considered here is a single
quantum well, consisting of a DMS layer of $Cd_{1-x}Mn_xTe$ (width $L$)
inside a thick layer of $Cd_{1-y}Mg_yTe$. The concentrations, $x$ and $y$
are chosen in such way to make the non-magnetic layer work as a barrier for
electrons and holes. The energy gap dependence on the $Mn$ concentration $x$%
, and on the $Mg$ concentration $y$, are given by \cite{dms1,euns,lebi}: 
\begin{equation}
E_g\,(300\,\,K)=(1.528+1.316x+1.316y)\,\,eV  \label{cdteg1}
\end{equation}
\begin{equation}
E_g\,(10\ K)=(1.595+1.607x+1.592y)\ eV  \label{cdteg2}
\end{equation}
\begin{equation}
E_g\,(4.2\,\,K)=(1.606+1.592x+1.592y)\,\,eV  \label{cdteg3}
\end{equation}
Since the problem of a single well has been extensively discussed in the
literature -- the only particularity of the present calculation lying on the
fact of the alloys taking the place of the pure binary compounds -- for the
sake of brevity, we will concentrate on the case of the double well,
assuming that the interested reader can map it into the case of a single
well easily \cite{mori}.

In this double quantum well structure, the $z$-direction phonon potential $%
\phi _{\vec{q}}(z)$ can be classified into CdTe confined modes, CdMnTe
confined modes, CdMnTe half space modes, and interface modes. The former
three kinds of phonon modes are the same as those in binary/binary double
quantum wells \cite{wang1}, and will not be discussed here in detail, either.

\subsection{The Interface phonon modes}

There are six symmetric, and six anti-symmetric interface modes in a
binary/ternary double quantum well. Their electric potentials $\Phi ({\bf r}%
) $ have similar expressions:

\subsubsection{Symmetric modes:}

\begin{equation}
\phi _{\vec{q}s}^{IF}(z)=\left\{ 
\begin{array}{l}
D_s\cosh qz \\ 
B_se^{-q\left| z\right| }+C_se^{q\left| z\right| } \\ 
A_se^{-q\left| z\right| }
\end{array}
\left. 
\begin{array}{l}
\text{ for }\left| z\right| <d/2 \\ 
\text{ for }d/2<\left| z\right| <d/2+L \\ 
\text{ for }d/2+L<\left| z\right|
\end{array}
\right. \right.  \label{modes}
\end{equation}
The frequencies of the symmetric interface modes are determined by the
relation:

\begin{eqnarray}
&&\varepsilon _1(\omega )\left\{ \varepsilon _1(\omega )(1+e^{-qd})\tanh
\left( qL\right) +\varepsilon _2(\omega )\right\} +  \label{disps} \\
&&\varepsilon _2(\omega )\left\{ \varepsilon _2(\omega )(1-e^{-qd})\tanh
\left( qL\right) +\varepsilon _1(\omega )\right\} =0  \nonumber
\end{eqnarray}
with

\begin{equation}
A_s=D_se^{q(d/2+L)}[\cosh qL\cosh \frac 12qd +\frac{\varepsilon _2(\omega )}{%
\varepsilon _1(\omega )} \sinh qL\sinh \frac 12qd],
\end{equation}

\begin{equation}
B_s=\frac 12D_se^{qd/2}[\cosh \frac 12qd-\frac{\varepsilon _2(\omega )} {%
\varepsilon _1(\omega )}\sinh \frac 12qd],
\end{equation}
and

\begin{equation}
C_s=\frac 12D_se^{-qd/2}[\cosh \frac 12qd+\frac{\varepsilon _2(\omega )} {%
\varepsilon _1(\omega )}\sinh \frac 12qd],
\end{equation}
where $D_s$ is determined by the normalization relation:

\begin{equation}
S\frac{\partial \varepsilon _2}{\partial \omega }q[2A_s^2e^{-q(2L+d)}
+D_s^2\sinh qd]+2S\frac{\partial \varepsilon _1}{\partial \omega}
q[B_s^2e^{-qd}(1-e^{-2qL})+C_s^2e^{qd}(e^{2qL}-1)] =\frac{\hbar e^2}{%
\varepsilon _0}
\end{equation}

\subsubsection{Anti-symmetric modes:}

\begin{equation}
\phi _{\vec{q}a}^{IF}(z)=\left\{ 
\begin{array}{l}
D_a\sinh qz \\ 
\pm B_ae^{-q\left| z\right| }\pm C_ae^{q\left| z\right| } \\ 
\pm A_ae^{-q\left| z\right| }
\end{array}
\left. 
\begin{array}{l}
\text{ for }\left| z\right| <d/2 \\ 
\text{ for }d/2<\left| z\right| <L+d/2 \\ 
\text{ for }L+d/2<\left| z\right|
\end{array}
\right. \right. ,  \label{modea}
\end{equation}
where the positive sign (+) is used for the range $z>0$ and the negative
sign (-) for the range $z<0$. The frequencies of the anti-symmetric
interface modes are determined by the relation:

\begin{eqnarray}
&&\varepsilon _1(\omega )\left\{ \varepsilon _1(\omega )(1-e^{-qd})\tanh
\left( qL\right) +\varepsilon _2(\omega )\right\} +  \label{dispa} \\
&&\varepsilon _2(\omega )\left\{ \varepsilon _2(\omega )(1+e^{-qd})\tanh
\left( qL\right) +\varepsilon _1(\omega )\right\} =0  \nonumber
\end{eqnarray}
with 
\begin{equation}
A_a=D_ae^{qd/2+qL}[\cosh qL\sinh \frac 12qd+\frac{\varepsilon _2(\omega )} {%
\varepsilon _1(\omega )}\sinh qL\cosh \frac 12qd],
\end{equation}
\begin{equation}
B_a=-\frac 12D_ae^{qd/2}[\frac{\varepsilon _2(\omega )}{\varepsilon
_1(\omega )}\cosh \frac 12qd-\sinh \frac 12qd],
\end{equation}
and 
\begin{equation}
C_a=\frac 12D_ae^{-qd/2}[\frac{\varepsilon _2(\omega )}{\varepsilon
_1(\omega )}\cosh \frac 12qd+\sinh \frac 12qd],
\end{equation}
where $D_a$ is determined by the normalization relation:

\begin{equation}
S\frac{\partial \varepsilon _2}{\partial \omega }q[2A_a^2e^{-q(2L+d)}+D_a^2%
\sinh qd+2S\frac{\partial \varepsilon _1}{\partial \omega }%
q[B_a^2e^{-qd}(1-e^{-2qL})+C_a^2e^{qd}(e^{2qL}-1)]=\frac{\hbar e^2}{%
\varepsilon _0}
\end{equation}

\subsection{Scattering rates}

With the electron (hole) wave functions $\psi _{n,\vec{k}}(z)$, and the
phonon electric potentials $\phi _{\lambda ,{\vec{q}}}({\bf r})$, we can
calculate the Fr\"{o}hlich interaction

\begin{equation}
H_{e-p}=\sum_{n,n^{\prime },\sigma }\sum_{\vec{k},\vec{q}}M(n,n^{\prime },%
\vec{q},q_z,\lambda )c_{n^{\prime },\vec{k}+\vec{q},\sigma }^{\dagger }c_{n,%
\vec{k},\sigma }A_\lambda (\vec{q}),
\end{equation}
where $A_\lambda (\vec{q})=b_{\vec{q},\lambda }+b_{-\vec{q},\lambda
}^{\dagger }$ is the phonon operator, and $M(n,n^{\prime },\vec{q}%
,q_z,\lambda )$ is the Fr\"{o}hlich interaction matrix in the structure: 
\begin{equation}
M(n,n^{\prime },\vec{q},\lambda )=-e\int dz\psi _{n^{\prime }}^{*}(z)\phi
_{\lambda ,{\vec{q}}},(z)\psi _n(z).
\end{equation}
Here, $n$ and $n^{\prime }$ are the indices of the electron subbands, and $%
\lambda $ denotes the phonon branch, this latter including all other indices
except the 2-D wavevector $\vec{q}$.

The zero-temperature scattering rate (inverse of the transition lifetime)
for an electron in an initial state $(n,\vec{k})$ into a subband $n^{\prime
} $ by emission of an optical phonon {\it {via}} the Fr\"{o}hlich
interaction, becomes:

\begin{eqnarray}
\frac 1{\tau (n,n^{\prime },\vec{k})} &=&W(n,n^{\prime },\vec{k})=2\pi \sum_{%
{\vec{k}}^{\prime },{\vec{q}}}\delta (E_f-E_i)\left| \left\langle n^{\prime
},\vec{k}^{\prime },1\left| H_{e-p}\right| n,\vec{k},0\right\rangle \right|
^2 \\
&=&\frac 1{2\pi }\sum_\lambda \int_0^\infty k^{\prime }\,dk^{\prime
}\int_0^{2\pi }d\theta \delta (\frac{k^{\prime 2}}{2m_t^{*}}-\frac{k^2}{%
2m_t^{*}}-\omega _\lambda )\left| M(n,n^{\prime },\vec{q}_0,\lambda )\right|
^2,  \nonumber
\end{eqnarray}
where the summation is extended to all final unoccupied states in the
subband $n^{\prime }$. Here $\vec{k}$ ($\vec{k}^{\prime }$) is the 2D
wavevector of the initial (final) state, $\vec{q}_0=\vec{k}^{\prime }-\vec{k}
$ , and $\theta $ is the angle between $\vec{k}$ and $\vec{k}^{\prime }$. $%
m_t^{*}$ is the average transversal ($x-y$ plane) effective mass of the
carrier \cite{andre}.

\section{Results}

The modified random-element isodisplacement ($MREI$) model \cite
{pete,euns,genz} is employed to obtain the bulk vibrational modes in $%
Cd_{1-x}Mn_xTe$ and $Cd_{1-y}Mg_yTe$, with parameters given by Eunson Oh et
al \cite{euns}. The corresponding dielectric functions {\it {versus}} the
phonon energy for $CdTe$, $Cd_{0.5}Mn_{0.5}Te$ and $Cd_{0.1}Mg_{0.9}Te$ are
shown in Fig. \ref{fig02}, according to Eqs. (\ref{diel1}) and (\ref{diel2}%
). In the following discussion, we use $\omega _{\text{T1}}$ (17.5 meV), and 
$\omega _{\text{L1}}$ (20.6 meV) to denote the phonon modes in bulk $CdTe$, $%
\omega _{\text{T2C}}$, $\omega _{\text{L2C}}$, $\omega _{\text{T2M}}$, and $%
\omega _{\text{L2M}}$ to denote the phonon modes in bulk $CdMnTe$, and $%
\omega _{\text{T3C}}$, $\omega _{\text{L3C}}$, $\omega _{\text{T3M}}$, and $%
\omega _{\text{L3M}}$ to denote the phonon modes in bulk $CdMgTe$, in
sequence from lower to higher energy. Eq. (\ref{cdteg2}) is used to estimate
the barrier potential for electrons ($70\%$ of the energy gap mismatch) and
holes ($30\%$ of the energy gap mismatch). The lowest two subbands for
either electrons and holes are included, when calculating the scattering
rates.

We consider, at first, the case of the $CdTe/CdMnTe$ double quantum well
with $L=50$ \AA\ and $d=10$ \AA . In Fig. \ref{fig03}, we show the energy
spectrum of the interface modes in $CdTe/Cd_{1-x}Mn_xTe$ double well with $%
x=0.1$, $0.5$, and $0.9$. Bulk samples of $Cd_{1-x}Mn_xTe$ have the
zinc-blende structure only for $x\leq 0.7$. In MBE grown samples, however,
the zinc-blend structure can be achieved even for $x=1$. Eqs.(\ref{disps})
and (\ref{dispa}) have solutions in the range of $\varepsilon _1/\varepsilon
_2\in [-1,0)$ and $(-\infty ,-1]$. Differently from a binary/binary double
well, in a binary-ternary structure there are six symmetric, and six
anti-symmetric interface modes, appearing in the ranges $[\omega _{\text{T1}%
},\omega _{\text{T2C}}]$, $[\omega _{\text{L2C}},\omega _{\text{L1}}]$, and $%
[\omega _{\text{T2M}},\omega _{\text{L2M}}]$, respectively. At low Mn
concentrations, $\omega _{\text{T2C}}$, and $\omega _{\text{L2C}}$ are close
to the phonon modes of $CdTe$, i.e., $\omega _{\text{T1}}$ and $\omega _{%
\text{L1}}$. On the other hand, $\omega _{\text{T2M}}$, and $\omega _{\text{%
L2M}}$ coincide with the localized mode associated to the $Mn$ impurity in $%
CdTe$. For $x=0.1$, we observe a small dispersion in the interface modes. In
the other limit, when the concentration of $Mn$ approaches $x=1.0$, the
frequencies of the phonon modes in bulk $Cd_{1-x}Mn_xTe$ approach the
frequencies of the pure bulk $MnTe$, plus that of the $Cd$ impurity mode.
This results in the increase of the dispersion of the 3rd--6th (in order of
increasing energy) interface modes. The dispersion of the 1st and the 2nd
interface modes are small in all range of concentration. This happens
because the frequency of the $Cd$ impurity mode in $MnTe$ is almost the same
as that of the transverse mode in $CdTe$.

In Fig. \ref{fig04}, we show the intra-subband and the inter-subband
electron and hole scattering rates due to the emission of $CdTe$ confined
phonon modes (grouped inside the circle C), as well as interface modes, as
functions of the initial electron kinetic energy for $x=0.1$, $0.5$, and $0.9
$. The solid lines, the dashed lines, the dotted lines and the chained lines
are the results corresponding to the transitions involving subbands $%
1\rightarrow 1$, $2\rightarrow 2$, $1\rightarrow 2$, and $2\rightarrow 1$,
respectively. We observe a sudden change in the electron scattering rates
due to the confined phonons as $x$ goes from 0.1 to 0.5. A saturation occurs
after that concentration. The scattering rates for holes (due to confined
phonons), on the other hand, are stronger than that for electrons (by a
factor of 3, approximately, in large concentrations), and, differently from
the electrons, are weakly dependent on $x$. The dependence of the scattering
rates due to the interface modes on the magnetic ions concentration is, on
the other hand, much weaker. Our results show, considering the order from
low to high energy in Fig. \ref{fig03}, that only the 3rd to the 6th modes
give significant scattering rates, both for electrons and holes. Only the
symmetric modes contribute to the intra-subband transitions and only the
anti-symmetric modes contribute to the inter-subband transitions. Our
calculation, performed for each individual mode, shows also that, in the
case of $x=0.1$, the third symmetric interface mode dominates the
intra-subband scattering rate, and the forth anti-symmetric interface mode
dominates the inter-subband scattering rate. The 3rd, and the 6th symmetric
modes generate almost the same scattering rates for the transition $%
1\rightarrow 1$, and $2\rightarrow 2$, but the 4th and the 5th symmetric
modes give scattering rates for the transition $1\rightarrow 1$, four times
bigger than that for the transition $2\rightarrow 2$. The scattering rates
for the transitions $1\rightarrow 2$ and $2\rightarrow 1$, due to
anti-symmetric interface modes, on the other hand, have almost the same
values at definite initial kinetic energy. With the increase of the $Mn$
concentration, the 5th and the 6th modes become more and more important. At $%
x=0.5$, the 3rd, 4th, and 6th symmetric modes give important contributions
for the intra-subband scattering. The same occurs for the 4th, 5th, and the
6th anti-symmetric modes, for the inter-subband scattering. In the case of
holes, the most remarkable difference with the electron scattering rates is
that the 4th interface phonon mode gives a greatly enhanced contribution.
Besides, the hole scattering rate due to the 3rd and the 6th interface modes
show an obvious increase with the increase of hole energy, for intra-subband
transitions. 
Next, we consider the electron (hole) phonon interaction in a
ternary/ternary $Cd_{1-x}Mn_xTe/Cd_{1-y}Mg_yTe$ single well, where the DMS
layer works as a well (after the proper choice of $x<y$). The phonon modes
in this structure can be obtained in the same way as that for a
binary/binary single well\cite{mori}, by just substituting the dielectric
function of the binary compound (Eq. \ref{diel1}) the dielectric function of
the ternary alloy (Eq. \ref{diel2}). The number of phonon modes in
ternary/ternary single well is twice that in a binary/binary single well. In
Fig. \ref{fig05}, the dispersion relations of the interface modes are shown
for $Cd_{1-x}Mn_xTe/Cd_{1-y}Mg_yTe$, with two different compositions:(i) $%
x=0.1$, $y=0.5$, and (ii) $x=0.5$, $y=0.9$. The symmetric modes (solid
lines) appears in the range $\varepsilon _2/\varepsilon _3\in (-\infty ,-1]$
and the anti-symmetric modes (dashed lines) in the range $\varepsilon
_2/\varepsilon _3\in [-1,0)$. Observing the dashed line ($CdMnTe$) and the
dotted line ($CdMgTe$) for dielectric function in Fig. \ref{fig02}, we
observe that four symmetric and four anti-symmetric interface modes appear
in the ranges $[\omega _{\text{T2C}},\omega _{\text{T3C}}]$, $[\omega _{%
\text{L3C}},\omega _{\text{L2C}}]$, $[\omega _{\text{T2M}},\omega _{\text{L2M%
}}]$ and $[\omega _{\text{T3M}},\omega _{\text{L3M}}]$.

The electron and hole scattering rates are shown in Fig. \ref{fig06}.
According to our calculation for individual modes, in the case (i) the $CdTe$%
-like modes give the major contribution among the confined modes. In the
case (ii), the $MnTe$-like modes dominate. The anti-symmetric interface
modes give scattering rates ten times smaller than the symmetric modes. The
contribution from the lowest energy interface mode is negligeable comparing
to other modes. 
\section{Comments}

Our calculation is performed at T=0, and we have only
considered the scattering assisted by the emission of optic modes. For that
reason, in some cases, the scattering rate is zero below a certain energy
threshold, corresponding to a minimum energy transfer equal to the energy of
the lowest lying relevant optic vibrational mode. For the same reason, a
saw-like structure may appear in the total scattering rates due to the
emission of interface modes. The results obtained by using the parameters of
the double well structure, shows no threshold for the $2\rightarrow 1$
transition for electrons, because the energy difference between the bottom
of the subbands is higher than the energy of the lowest lying relevant mode.
A remarkable result, which is a consequence of the widths chosen for the
wells and barrier, is that the interface modes dominate the electron
scattering in the whole range of the magnetic ion concentration, whereas the
confined modes dominate the low energy hole scattering in the intermediate
and high magnetic ion concentration.

Contrarily to the case of the double well, in the ternary/ternary single well
structure, for both choices of composition, there is just a weak dependence of
the transition rates on the concentrations $x$ and $y$. This is a
consequence of the fact in structures those structures the barrier height did
not change, because we took care of keeping the DMS material as the well.
The differences on the scattering rates reflects, in this case, the changes
on the optic phonon dispersion relations. As in the case of the double well
structure, the scattering by interface phonons dominate the intra-subband
electron transition. In what concerns the inter-subband, however, almost an
order of magnitude weaker, it is hard to distinguish between the
contributions due to the interfaces from those of the confined modes. We had
to use arrows indicating each mode, in that case. For holes, above an
initial kinetic energy just a few eV above the energy threshold, the
intra-subband scattering rates are dominated by the interface modes, as in
the case of electrons. However, the inter-subband transitions are dominated
by the confined phonons.

In summary, employing the $MREI$ model to calculate the optical phonon modes
in bulk ternary alloy $CdMnTe$ and $CdMgTe$, and, after that the dielectric
function of those material in the energy range of $\sim 10meV$, we studied
the confinement of optical phonon modes in binary/ternary double well and
ternary/ternary single well structures within the framework of dielectric
continuum model. Carrier scattering rates of emission of optical phonons in
those structures are calculated.

\acknowledgements 
ICCL and XFW are grateful to C. Testelin for very fruitful discussions. This
work was partially supported by FAPERJ and CNPq, in Brazil.

\begin{figure}[tbh]
\caption{Schematic drawing of the $CdTe/Cd_{1-x}Mn_xTe$ double well
structure.}
\label{fig01}
\end{figure}

\begin{figure}[tbh]
\caption{Dielectric functions of $CdTe$ (solid line), $Cd_{0.5}Mn_{0.5}Te$
(dashed line), and $Cd_{0.1}Mg_{0.9}Te$ (dotted line), {\it versus} phonon
energy.}
\label{fig02}
\end{figure}

\begin{figure}[tbh]
\caption{Energy spectra of the symmetric and anti-symmetric interface phonon
modes in a CdTe/Cd$_{1-x}$Mn$_x$Te double well with $x=0.1$, $x=0.5$ and $%
x=0.9$.}
\label{fig03}
\end{figure}

\begin{figure}[tbh]
\caption{Intra-subband (solid line for 1$\rightarrow $1 and dashed line for 2%
$\rightarrow $2) and inter-subband (dotted line for 1$\rightarrow $2 and
chained line for 2$\rightarrow $1) scattering rates of the electrons and
holes due to the emission of $CdTe$ confined phonons (grouped inside circle
C), and of interface modes, as functions of the initial kinetic energy. The
structure parameters are $L=50$ \AA\ and $d=10$ \AA , with $x=0.1$, $x=0.5$
and $x=0.9$.}
\label{fig04}
\end{figure}

\begin{figure}[tbh]
\caption{The symmetric (solid lines), and the anti-symmetric (dashed lines)
interface phonon dispersion curves in a $Cd_{1-x}Mn_xTe/Cd_{1-y}Mg_yTe$
single quantum well of width $L=50$ \AA , and (a) $x=0.1$, $y=0.5$, (b) $%
x=0.5$, $y=0.9$.}
\label{fig05}
\end{figure}

\begin{figure}[tbh]
\caption{Scattering rates due to emission of confined phonons (grouped
inside circle C) and interface modes, as functions of the initial kinetic
energy in the $Cd_{1-x}Mn_xTe/Cd_{1-y}Mg_yTe$ single well.}
\label{fig06}
\end{figure}

\end{document}